\documentclass[%
 reprint,
 amsmath,amssymb,
 aps,
]{revtex4-2}

\usepackage{soul}
\usepackage{graphicx}
\usepackage{dcolumn}
\usepackage{bm}
\usepackage[hidelinks]{hyperref}

\usepackage{color}
\newcommand{\dd}{\mathrm{d}}

\begin{document}

\preprint{APS/123-QED}

\title{De Sitter Complexity Grows Linearly in the Static Patch}

\author{\href{mailto:vyshnav.vijay.mohan@gmail.com}{Vyshnav Mohan}${}^{\spadesuit}$ and \href{mailto:watse.sybesma@su.se}{Watse Sybesma}${}^{\heartsuit}$}
 \affiliation{
${}^{\spadesuit}$Science Institute, University of Iceland, 
Dunhaga 3, 107 Reykjavík, Iceland
\\
${}^{\heartsuit}$Nordita, KTH Royal Institute of Technology and Stockholm University,
Hannes Alfv\'{e}ns v\"{a}g 12, 106 91 Stockholm, Sweden
 }
 
\date{\today}

\begin{abstract}
The observable universe has undergone periods of expansion that are well approximated by de Sitter (dS) space. Still lacking is a quantum mechanical description of dS, both globally and when restricted to the static patch. We develop a novel prescription for computing holographic complexity in the dS static patch to determine its microscopic features. Specifically, we propose that the natural candidate for dS complexity is the volume of extremal \emph{timelike} surfaces restricted to the static patch, anchored to the cosmological horizon or an observer worldline. 
Our anchoring prescription provides a clear definition of a reference state, overcoming a common ambiguity in prior definitions of de Sitter holographic complexity. The late-time growth of our complexity functional is linear and proportional to the number of degrees of freedom associated to the cosmological horizon, and therefore does not exhibit hyperfast growth. Our results imply the dS static patch is characterized by a quantum mechanical system, with a finite dimensional Hilbert space whose evolution is governed by a chaotic Hamiltonian. 

\end{abstract}

\maketitle

%

\section{Introduction}

Profound insights into the quantum nature of Anti-de Sitter (AdS) spacetime are provided through a powerful notion called holography, which in essence allows for mapping an AdS (quantum) gravitational theory to a dual quantum mechanical system \cite{Maldacena:1997re,Gubser:1998bc,Witten:1998qj}.
Meanwhile, it is expected that the expansion of our current and early inflationary universe is well-described by de Sitter (dS) spacetime, see e.g. \cite{baumann2022cosmology}. Realizing holography for dS could provide a window for studying quantum aspects of the universe.

AdS holography has given rise to a fruitful interplay between gravitational physics and quantum information theory, introducing geometric probes in spacetime that have a dual quantum information interpretation.
In particular, proposals of holographic complexity, initially inspired by notions of computational complexity in quantum circuits, have been employed to discern quantum features of spacetime, see e.g. \cite{Baiguera:2025dkc} for a recent review. 
The standard computation of black hole holographic complexity using the Complexity = Volume (CV) prescription involves spacelike extremal volume surfaces anchored on timelike surfaces \cite{Susskind:2014rva,Stanford:2014jda}. 
From such studies, a picture emerges where the late-time linear growth of holographic complexity is in conjunction with a dual chaotic dynamical evolution (see, for example, \cite{Barbon:2019wsy}). 
A tool like complexity could, on the one 
hand, reveal quantum aspects of dS, and on the other hand, help working towards understanding dS holography.

The main result we present in this letter is a new CV proposal applicable to dS spacetime that instead involves anchoring \textit{timelike} extremal volumes. Our refinement exhibits late-time linear growth and is confined within the horizons of a single static patch. Our results suggest that the static patch of de Sitter space could be described by a finite-dimensional quantum system with a chaotic Hamiltonian. This is consistent with earlier conjectures surrounding the finiteness of the de Sitter Hilbert space \cite{Fischler:1998st,Parikh:2004wh,Banks:2005bm}.

\begin{figure}[h!]
    \centering
      \includegraphics[width=.9\linewidth]{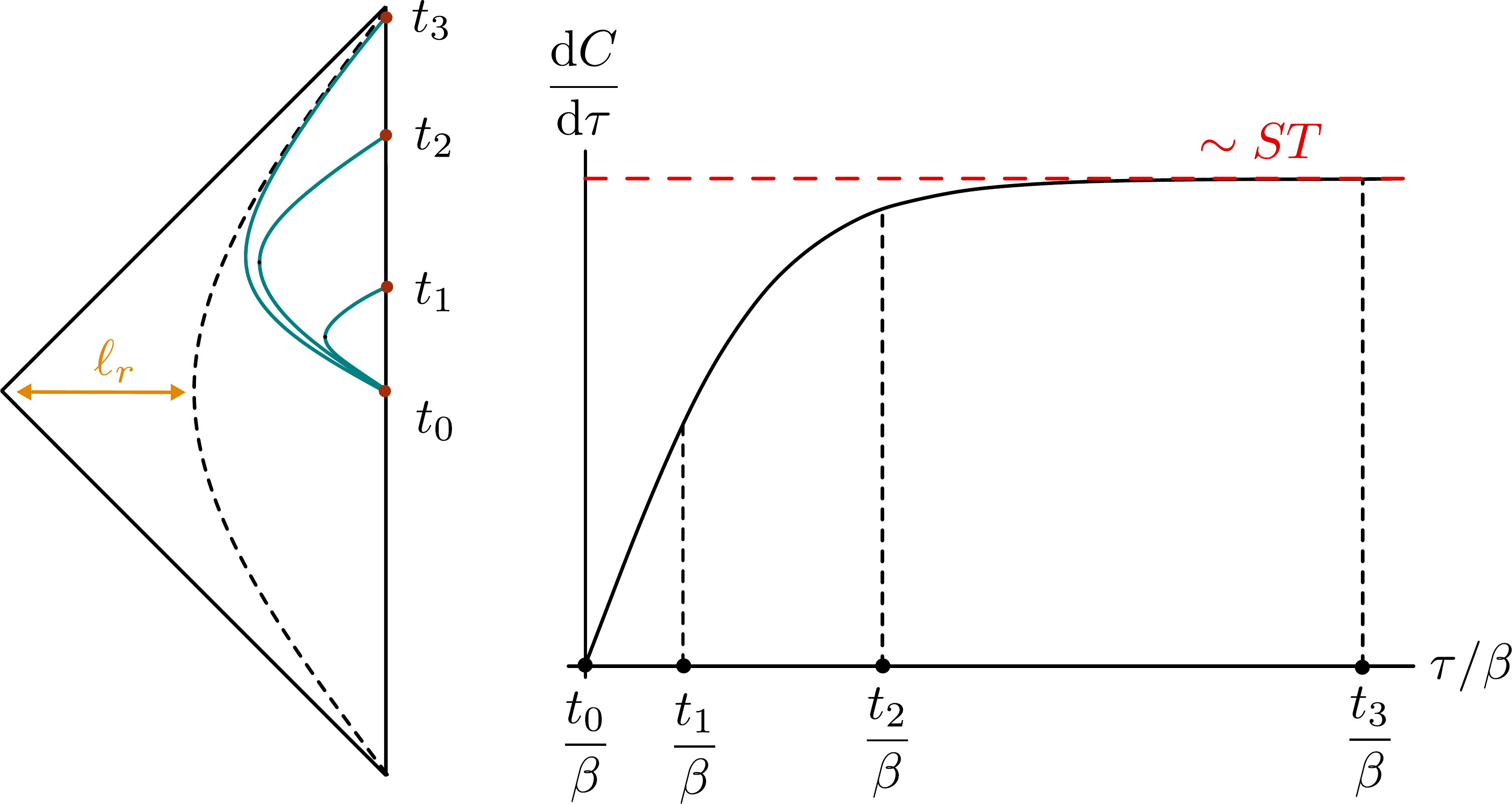}
    \caption{The left conformal diagram traces an observer on the vertical line advancing through the static patch, connecting with extremized surfaces to $t_{0}$ at different moments in time. Due to the time translation symmetry of the problem,  the precise choice of $t_0$ does not matter. The right graph shows the corresponding complexity growth. Here, $\beta=1/T$ is the inverse temperature and $S$ is the entropy both associated to the cosmological horizon. When $\tau \gg \beta$, the growth rate saturates.}
    \label{frontmatter1}
\end{figure}

Our prescription makes use of three key ingredients, which we illustrate in Figure \ref{frontmatter1}. (i) Explicit inclusion of an observer through anchoring, inspired by recent advances \cite{Chandrasekaran:2022cip,Maldacena:2024spf}, although we alternatively also consider horizon anchoring. In both cases, it holds that a state at a given moment $\tau=t_{0}$ can be interpreted as a reference state. (ii) Extremizing the volume $\mathcal{V}$ of a \emph{timelike} surface, we
point out that as time evolves, it will converge to an accumulation surface, allowing the late-time behavior to approach a linear growth. 
(iii) To define complexity 
\begin{equation}\label{eq:cvdef}
    C:= \frac{\mathcal{V}}{G_N \ell_{r}}\,,
\end{equation}
we compare the volume $\mathcal{V}$ to a relevant complex-valued reference scale $\ell_{r}$ (and Newton's constant $G_{N}$) that is determined by generalizing a proposal by \cite{Couch:2018phr} via a min-max procedure. 
Owing to the timelike nature of the surface, the volume picks up a complex phase factor of $i$ that seems troublesome for interpreting a real complexity. However, our reference scale ensures the complexity to be real valued, by naturally counteracting the complex phase factor $i$ coming from the volume.

Importantly, our results evade the \textit{hyperfast} growth of complexity encountered in dS when considering spacelike extremal surfaces \cite{Susskind:2021esx}.
In the literature, this behavior has also been ameliorated by introducing a cutoff surface near future infinity, resulting in a linear growth that depends on the specific details of the cutoff surface, which however make the result more difficult to interpret \cite{Jorstad:2022mls}.
In the work by \cite{Aguilar-Gutierrez:2023zqm}, it was revealed that a family of surfaces exists that bypasses this hyperfast growth, allowing for \emph{various} possible late-time growth behaviors, but not exclusively linear.

In the next section, we will explicitly compute the complexity growth via \eqref{eq:cvdef} through `observer anchoring'.


\section{Complexity Through Observer Anchoring}
%

%
\subsection{Setting up the timelike surface}
\begin{figure}[h!]
  \centering
\includegraphics[width=0.75\linewidth]{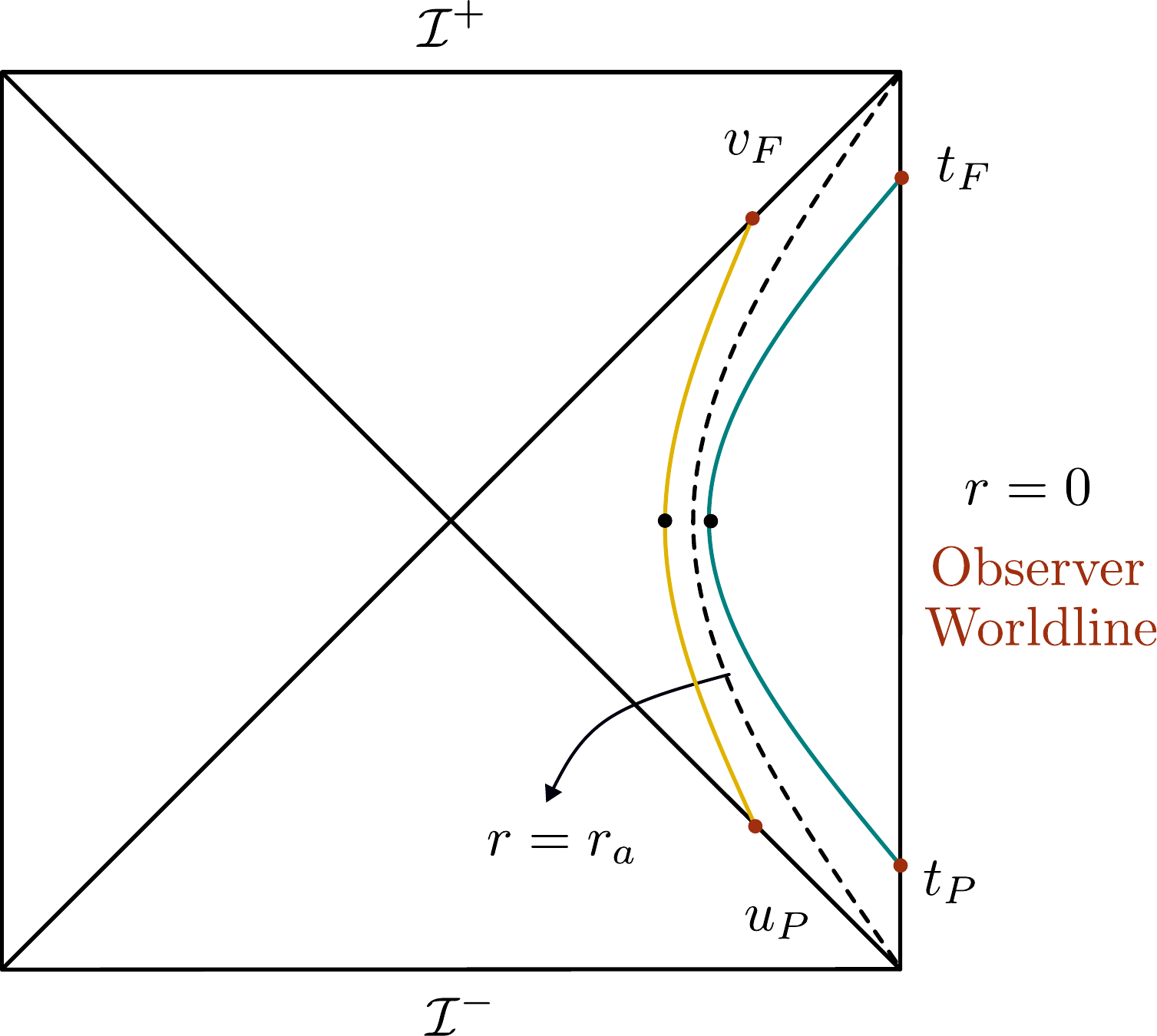}
  \caption{The blue curve corresponds to a timelike extremal surface anchored to the observer worldline while the yellow curve corresponds to an extremal surface anchored to the cosmological horizon. The turning points of the extremal surfaces are shown by black dots. At late anchoring times, the surfaces approach the accumulation surface, shown with dashed lines.}
  \label{dspenrosefig}
\end{figure}
In static coordinates, a $d+1$-dimensional de Sitter metric is given by
\begin{equation}
    \dd s^2 = -f\dd t^2 + \frac{\dd r^2}{f} + r^2 \dd \Omega_{d-1}^2\,, \ \text{where} \ f = 1-\frac{r^2}{L^2}\,,\label{dsmetric}
\end{equation}
and where $\dd \Omega_{d-1}^2$ refers to a transverse $(d-1)$-sphere.
The cosmological horizon is located at $r=L$. We restrict our attention to $d \geq 2$ and will comment on the $d=1$ case in the discussion section. Consider a spherically symmetric timelike codimension-one surface in the right static patch \footnote{Timelike codimension-two surfaces in de Sitter space have been previously studied, and the corresponding boundary dual has been proposed to be what is referred to as pseudo-entropy (see, for example, \cite{Doi:2022iyj,Doi:2023zaf,Narayan:2022afv,Narayan:2023ebn}).}. We anchor the surface to the $r=0$ antipodal surface, as shown in Figure \ref{dspenrosefig}. For simplicity, but without any loss of generality, we will work with the time-symmetric case where the surface is anchored at times $t_F=-t_P=\tau/2$. The (to be) extremized surface starts off at the observer worldline, goes through the static patch, reaches a turning point before coming back to the $r=0$ surface. Due to the symmetry of the problem the turning point lies on the $t=0$ surface. Since the growth rates of the volume will ultimately depend only on the time difference between the anchoring points, we can also choose $t_P=0$ and set $t_F=\tau$, which is what we are ultimately interested in and is depicted in Figure \ref{frontmatter1}.

Now let us extremize the volume functional of these surfaces and compute their growth rate. The general procedure follows the spacelike case of \cite{Carmi:2017jqz}, to which we refer the reader for details; here we focus instead on the crucial differences that arise for a timelike surface. It is convenient to work with the Eddington-Finkelstein coordinate $v$ defined through the relation
\begin{equation}
   v= t+r^{*}\,, \quad  r^{*}(r) =\int_0^r \frac{\dd r^{\prime}}{f(r^{\prime})}\,.
\end{equation}
The volume of the timelike surface can then be written as the following integral 
\begin{equation}
    \mathcal{V} = \Omega_{d-1}\int \dd\lambda \ r^{d-1}\sqrt{-f(r) \dot{v}^2+2\dot{v}\dot{r}}\,, \label{volumeintegral}
\end{equation}
where we have parametrized the surface as $\left(v(\lambda), r(\lambda)\right)$. We have also used $\Omega_{d-1}$ to denote the volume of a $(d-1)$-dimensional unit sphere. The crucial difference from the spacelike case is that the volume $\mathcal{V}$ in \eqref{volumeintegral} is now imaginary.

Since the volume functional does not depend explicitly on $v$, there exists a conserved quantity which we denote by $p$. The volume functional is also reparametrization invariant, allowing us to choose the parameter $\lambda$ as follows
\begin{equation}
    r^{2d-2} \left(-f \dot{v}^2+2 \dot{v} \dot{r}\right)=-1\,. \label{reparametrizationeq}
\end{equation}
The sign on the right-hand side is negative because we are considering timelike surfaces. The conservation equation and the choice of the affine parameter can then be used to arrive at the following equations of motion
\begin{eqnarray}
p &=r^{2(d-1)}(f(r) \dot{v}-\dot{r})\,, \label{eom1}\\
r^{2(d-1)} \dot{r}^2 &=-f(r)+r^{-2(d-1)} p^2\,.\label{eom2}
\end{eqnarray}
We will analyze these equations of motion in the remainder of this section.

\subsection{Extremizing volume $\mathcal{V}$}
The surface has a turning point, whose location we denote by $r=r_{\min}$. The location of the turning point can be obtained by setting $\dot{r}$ to zero:
\begin{equation}
    r_{\min}^{2(d-1)}f(r_{\min})= p^2 \label{turningpointeq}
\end{equation}
We choose $\lambda$ to increase in the upward direction in the Penrose diagram in Figure \ref{dspenrosefig}. As a result, $\dot{v}>0$ when $r=r_{\min}$. Evaluating \eqref{eom1} at the turning point, we find that $p$ is positive.

Using the coordinate transformation, 
\begin{equation}
    \dd\tilde{r} = r^{2d-2} \dd r\,,
\end{equation}
we can rewrite \eqref{eom2} as a particle scattering off an effective potential \cite{Carmi:2017jqz,Gautason:2025ryg}:
\begin{equation}
    \dot{\tilde{r}}^2 + V_{\text{eff}} = p^2\,, \qquad \text{where} \ V_{\text{eff}}(r) = r^{2d-2}f(r)\,.
\end{equation}
\begin{figure}
  \centering
    \includegraphics[width=0.85\linewidth]{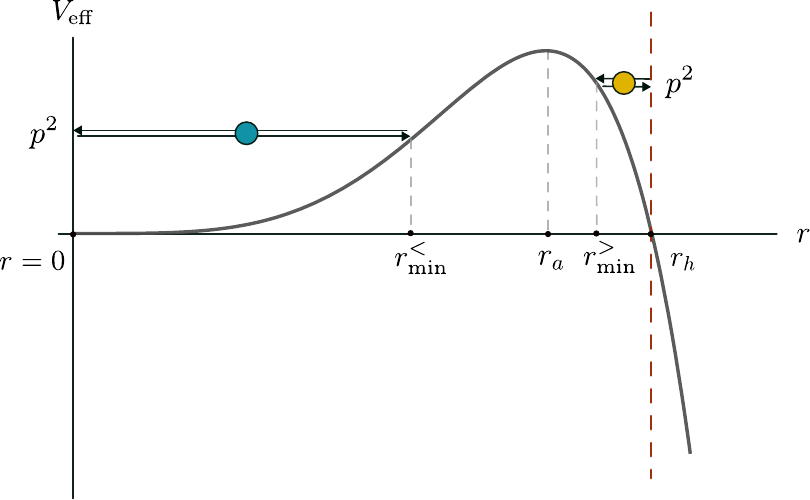}
  \caption{The scattering picture of the extremal surface. The surfaces in Figure \ref{dspenrosefig} can be thought of as a particle with total energy $p^2$, scattering off an effective potential $V_{\text{eff}}$. The extremal surfaces anchored to the antipodal surface (blue) scatter off the left side of the potential, while those anchored to the horizon (yellow) scatter off the right side. At late times, the turning points of the surfaces, labelled by $r_{\min}^{<}$ and $r_{\min}^{>}$, approach the top of the potential, denoted by $r_a$.}
  \label{figures/effectivepotentialfig}
\end{figure}
Here, $p^2$ is the total energy of the particle.



The effective potential in the scattering picture
is useful in inferring the properties of the extremal surface without explicitly solving the equations of motion. We have plotted $V_{\text{eff}}$ as a function of $r$ in Figure \ref{figures/effectivepotentialfig}. The potential has a maximum at some $r=r_a$. 
We refer to this constant-$r$ surface as the \emph{accumulation surface}. The location of the accumulation surface is given by
\begin{equation}
    \left.\frac{\dd}{\dd r}\left(r^{2d-2}f(r)\right)\right|_{r=r_a}=0\, \implies r_a = L \sqrt{\frac{d-1}{d}}\,.\label{accumulationequation}
\end{equation}
Here, we have ignored the trivial $r_a=0$ solution.

The accumulation surface plays an important role in our computation, as the extremal surface approaches this constant-$r$ surface in the infinite $\tau$ limit. A quick way to see this is by using the reasoning from \cite{Stanford:2014jda}. Namely, in this limit, the extremal surface should become time-translation invariant and, therefore, must lie at constant $r$. Setting $\dot{r} = 0$ in \eqref{volumeintegral}, we find that the extremization condition reduces to \eqref{accumulationequation}. As a result, the infinite $\tau$ extremal surface coincides with the accumulation surface.

The equations of motion \eqref{eom1} and \eqref{eom2} also admit a trivial solution in which the extremal surface is always located at $r = 0$. The volume of this surface is identically zero, and the trivial solution of \eqref{accumulationequation} corresponds to the infinite $\tau$ limit of this class of solutions. We will not consider these solutions or their associated accumulation surface due to their triviality \footnote{As this trivial solution will turn out to be absent when anchoring to the horizon, one could muse that this solution can be associated to some sort of `internal' complexity of the observer.}.

The energy of the particle, $p^2$, is a monotonic function of $\tau$ (see \eqref{tauofpeq} for the explicit relation; numerical evaluation of the integral can be used to confirm the monotonic behavior). Using the fact that the extremal surface asymptotes to the accumulation surface, we can use \eqref{turningpointeq} to see that in the $\tau \to \infty$ limit, $p^2$ approaches $V_{\text{eff}}(r_a)$~\cite{Carmi:2017jqz,Gautason:2025ryg}. In the scattering picture, this simply corresponds to the particle approaching the top of the potential.

Using \eqref{reparametrizationeq}, we can rewrite the volume integral as follows
\begin{equation}
\mathcal{V} = 2i\Omega_{d-1}\int_{\lambda(r=r_{\min})}^{\lambda(r=0)} \dd \lambda = 2i\Omega_{d-1}\int_{r_{\min}}^{0} \frac{\dd r}{ \dot{r}}\,.
\end{equation}
We have used the symmetry about the $t=0$ surface to simplify the expression. We have also chosen the positive branch for the square root of $-1$ for $i$. 
The negative branch could also be used, provided the same choice is made when computing the reference scale $\ell_{r}$ later in the section. 
The order of the integration limits follows from our choice of $\lambda$ increasing in the upward direction. Since $\dot{r}$ is negative in the region of integration, we can plug in the equation of motion \eqref{eom2} to get
\begin{equation}
    \mathcal{V} = 2i\Omega_{d-1} \int^{r_{\min}}_{0} \dd r \frac{r^{2(d-1)}}{\sqrt{-f(r) r^{2(d-1)}+p^2}}\,.
\end{equation}
The equation of motion for $\dot{v}$ can be re-expressed as the following integral
\begin{equation}
    \int \dd v = \int  \frac{\dd r}{f(r)}\left[\frac{p}{r^{2(d-1)}\dot{r}}+1\right]\,.
\end{equation}
Choosing appropriate ranges of integration and the associated sign of $\dot{r}$, we get
\begin{equation}
    r^{*}_{\min}+\frac{\tau}{2}
    =  
    \int^{r_{\min}}_{0} \frac{\dd r}{f(r)}\left[\frac{p}{ \sqrt{-f(r) r^{2(d-1)}+p^2}}+1\right]\,.\label{tauofpeq}
\end{equation}
The above expression implicitly relates the anchoring time $\tau$ to $p$ since $r_{\min}$ depends only on $p$.

To compute the growth rate, we employ a simple trick of rewriting the volume integral in the following form \cite{Carmi:2017jqz}
\begin{equation}
    \begin{aligned}
    -\frac{\mathcal{V}}{2 i \Omega_{d-1}}
    =
    \int^{r_{\min }}_{0} 
    \frac{\dd r}{f(r)}& \left[\sqrt{-f(r) r^{2(d-1)}+p^2}+p\right]\\
    &\hspace{2cm}-p\left(r^{*}_{\min}+\frac{\tau}{2}\right)\,.\\
    \end{aligned}
\end{equation}
Taking a derivative with respect to $\tau$ of the equation above, we obtain the simple expression
\begin{equation}\label{eq:growth1}
    \frac{\dd\mathcal{V}}{\dd \tau} = i  \Omega_{d-1} p  = i \Omega_{d-1} r_{\min}^{d-1}\sqrt{f(r_{\min})}\,. 
\end{equation}
At late times, $r_{\min}$ approaches $r_a$, and as a result, the growth rate of the timelike extremal surfaces can be approximated by a constant.

\subsection{Determining reference scale $\ell_{r}$}

Now that we have computed the volume $\mathcal{V}$, we need to determine the reference scale to compare it to, in order to evaluate complexity through \eqref{eq:cvdef}.
In the AdS black hole, for example, one typically chooses this scale to be the AdS length scale while in a Schwarzschild black hole, one chooses it to be the horizon radius. In \cite{Couch:2018phr}, it was suggested that this length scale can be universally defined as the maximum time to fall from the horizon to the accumulation surface. Here, we generalize the prescription to a min-max procedure. We begin by fixing the non-compact directions $r$ and $t$, and minimizing the proper time over all infalling curves. We then vary $r$ and $t$ to extremize the resulting proper time.

We can easily verify that this prescription reproduces all the length scales computed in \cite{Couch:2018phr}. Now let us use this prescription to compute the reference scale of dS. Since the accumulation surface is timelike, the `infalling' trajectories are spacelike. As a result, the `time' taken to `fall' from the horizon becomes imaginary. For a fixed $r$ and $t$ profile, the minimum proper length is given by a curve located at fixed angular coordinates. Now varying the $r$ and $t$ coordinates, we find that the maximal (imaginary) proper time is given by the $t=0$ surface. Note that this is similar to what we find in the black hole case, but the crucial difference is that the $t=0$ is timelike in the interior of a black hole.

The reference scale $\ell_{r}= \int \sqrt{-g_{\mu \nu} \dot{x}^{\mu}\dot{x}^{\nu}}$ is then given by

\begin{equation}\label{eq:referencescale}
    \begin{aligned}
        \ell_{r} 
        =i \int_{r_a}^{L} \frac{\dd r}{\sqrt{f(r)}} = i L \cos ^{-1}\sqrt{\frac{d-1}{d}}\,.
    \end{aligned}
\end{equation}

\subsection{Putting it all together}

We can now determine the growth rate of complexity by using its definition \eqref{eq:cvdef} and plugging in \eqref{eq:growth1} and \eqref{eq:referencescale}:
\begin{equation}\label{complexitygrwotheq}
    \frac{\dd C}{\dd \tau} = \frac{8 \pi \left(\frac{d-1}{d}\right)^{\frac{d-1}{2}}}{\sqrt{d}\cos ^{-1}\sqrt{\frac{d-1}{d}}} S T \propto S T \,,
\end{equation}
where $S$ and $T$ are the entropy and temperature of dS
\begin{equation}
    S = \frac{\Omega_{d-1} L^{d-1}}{4 G_N}\,, \quad\ \ T =\frac{1}{2\pi L}\,.
\end{equation}
Therefore, timelike complexity grows linearly at late times, with a growth rate proportional to the number of degrees of freedom, as expected for a finite-dimensional chaotic quantum system! 

Now let us look at the case where we fix $t_P=0$ and choose $t_F=\tau$ (see Figure \ref{frontmatter1}). Owing to the symmetry of the problem, the turning point is now located at $t=\tau/2$ and $r=r_{\min}$. Since the equations of motion are the same, we can proceed as in the previous case to find the same growth rate as in \eqref{complexitygrwotheq}.

%
\section{Complexity Through Horizon Anchoring}
%

Instead of anchoring the extremal surface onto the observer worldline, we can anchor it at the horizon, or in fact anywhere beyond $r_{a}$. To do this, we first define the other Eddington-Finkelstein coordinate $u=t-r^{*}$. The anchor points can now be specified in terms of the Eddington-Finkelstein coordinates $u$ and $v$ (see Figure \ref{dspenrosefig}). We will choose to anchor the surface at symmetric points by setting $v_F=-u_P=v/2$.

The equations of motion are once again given by \eqref{eom1} and \eqref{eom2}. The only difference in this setup is that the particle in the scattering picture starts at the horizon and scatters off the potential from the right side, as shown in Figure \ref{figures/effectivepotentialfig}. The volume integral now takes the form
\begin{equation}
    \mathcal{V} = 2i\Omega_{d-1} \int_{r_{\min}}^{L} \dd r \frac{r^{2(d-1)}}{\sqrt{-f(r) r^{2(d-1)}+p^2}}\,.
\end{equation}
Using the equation of motion for $\dot{v}$, we find that:
\begin{equation}
    \begin{aligned}
     \frac{v}{2}-r^{*}_{\min}
    &= \int_{v(r=r_{\min})}^{v(r=L)} \dd v \\
    &=  \int_{r_{\min}}^{L} 
    \frac{\dd r}{f(r)}\left[\frac{p}{ \sqrt{-f(r) r^{2(d-1)}+p^2}}+1\right].
    \end{aligned}
\end{equation}
This gives us
\begin{equation}
    \begin{aligned}
    -\frac{\mathcal{V}}{2 i \Omega_{d-1}}=\int^{r_{h}}_{r_{\min}} \frac{\dd r}{f(r)}& \left[
    \sqrt{-f(r) r^{2(d-1)}+p^2}+p\right]\\
    &\hspace{2cm}-p\left(\frac{v}{2}-r^{*}_{\min}\right)\,.\\
    \end{aligned}
\end{equation}
Taking a $v$-derivative, we get the integral
\begin{equation}
    \frac{\dd\mathcal{V}}{\dd v} = i  p  \Omega_{d-1} =i \Omega_{d-1} r_{\min}^{d-1}\sqrt{f(r_{\min})}\,. 
\end{equation}
At late times, $r_{\min}$ approaches $r_a$. This gives us the late-time growth rate
\begin{equation}
    \frac{\dd\mathcal{V}}{\dd v} \simeq  i \Omega_{d-1} r^{d-1}_a \sqrt{f(r_a)} \,. 
\end{equation}
Using the infalling time we have calculated in the previous section, we find the complexity growth to be given by
\begin{equation}
    \frac{\dd C}{\dd v} \simeq \frac{ i \Omega_{d-1}}{G_N \ell_{r}} r^{d-1}_a \sqrt{f(r_a)} \propto S T \,. 
\end{equation}
The growth rate matches the final expression we obtained in the previous section, equation \eqref{complexitygrwotheq}. This suggests that the dS stretched horizon complexity is, in fact, the same as the complexity seen by the observer.

\section{Discussion}
In this letter, we have proposed a new definition of holographic complexity in de Sitter spacetime, based on computing the volumes of extremal timelike surfaces within a single static patch.
We find late-time linear behavior for the complexity growth, without any signs of hyperfast growth. We find the same results using both an observer as well as a horizon perspective, both perspectives give us a clear interpretation of the reference state, namely the state at `$\tau=t_0$'. Our results suggest that the static patch in de Sitter has a dual description in terms of a finite-dimensional quantum chaotic system.

The above picture is further supported by the following observation. If we naively apply the results of \cite{Gautason:2025ryg} to our setup, then the complexity should saturate at a timescale given by the exponential of the area of the accumulation surface. This corresponds to $e^{\mathcal{O}(S)}$, exactly as expected for a chaotic system.

In the future, to find more evidence to support the interpretation of de Sitter as a chaotic system, one could perform shockwave experiments or study the switchback effect to uncover out-of-equilibrium effects. In particular, sending in a shockwave and computing the delay in the onset of linear growth can be used to extract the scrambling time, as discussed in \cite{Anegawa:2023dad,Baiguera:2023tpt,Baiguera:2024xju} using the standard CV computations in dS.

One may also revisit other holographic complexity proposals, such as Complexity = Action \cite{Brown:2015bva} and Complexity = Anything \cite{Belin:2021bga,Belin:2022xmt}, as studied in \cite{Jorstad:2022mls,Aguilar-Gutierrez:2023zqm}, by anchoring the Wheeler-DeWitt patch and codimension-one volume surfaces to timelike separated points on the observer's worldline. We expect these computations to confirm further and strengthen the linear growth found in this work.

Finally, despite carefully defining our results to be only valid beyond two dimensions ($d=1$), we find that the final result allows continuation to two dimensions \eqref{complexitygrwotheq}, predicting late-time linear growth. 
Meanwhile, we point out that often two-dimensional models have a higher dimensional pedigree. For example, two-dimensional de Sitter can be obtained through reducing three-dimensional de Sitter or near-extremal limits of four-dimensional de Sitter black holes, see \cite{Sybesma:2020fxg,Aalsma:2021bit} for discussions. One can always `dress' the two-dimensional volume functional with an appropriate power of the dilaton to match the complexity of its higher dimensional parent, see e.g. \cite{Schneiderbauer:2019anh,Anegawa:2023wrk}.

The reason we emphasize the two-dimensional realisation of our results is that it provides an exciting bench mark for matching conjectured dual systems (see for example \cite{Susskind:2021esx,Susskind:2022bia,Susskind:2022dfz,Narovlansky:2023lfz,Rahman:2022jsf,Rahman:2023pgt,Verlinde:2024znh,Rahman:2022jsf}). These quantum systems can be completely solved using so-called \textit{chord diagrams}, whose intersection numbers map to the lengths of geodesics in the bulk geometry \cite{Blommaert:2024ymv}. Since these are precisely the elements required for a quantum-mechanical definition of holographic complexity, the Krylov spread complexity in these dual descriptions is expected to match our predictions, analogous to the cases studied in \cite{Rabinovici:2023yex,Heller:2024ldz}.

%
\begin{acknowledgments}
We thank S. Baiguera, P. Caputa, M. Heller, K. Narayan, D. Patramanis, T. Schuhmann, A. Svesko, and L. Thorlacius for discussions and feedback. VM is supported in part by the Icelandic Research Fund under grant 228952-053 and by a doctoral grant from The University of Iceland Science Park. The work of WS is supported by a Starting Grant 2023-03373 from the Swedish Research Council.
\end{acknowledgments}

\bibliography{apssamp}

\end{document}